\documentclass[journal=jacsat,manuscript=article,layout=twocolumn]{achemso}

\usepackage[version=4]{mhchem} 
\usepackage{placeins}
\usepackage[T1]{fontenc} 
\usepackage{lmodern}     

\setkeys{acs}{maxauthors=25}

\makeatletter
\let\acs@onecol@maketitle\maketitle
\let\maketitle\relax 
\renewenvironment{abstract}{\acs@collect\acs@onecol@abstract}{}
\newcommand\acs@onecol@abstract[1]{%
  \global\long\def\acs@onecol@abstract@text{%
    \par
    \section*{\abstractname}%
    #1%
    \acs@abstract@extras
    \par
    \vspace*{\acs@space@post@email}%
  }%
  \g@addto@macro\@maketitle{\acs@onecol@abstract@text}%
  \acs@onecol@maketitle
}
\makeatother
\author{Brenna C. Bierman}
\affiliation[University of Wisconsin-Madison Chemistry]
{Department of Chemistry, University of Wisconsin-Madison, Madison, Wisconsin 53706, United States}
\author{Taeren Yang}
\affiliation[University of Wisconsin-Madison MS\&E]
{Department of Materials Science and Engineering, University of Wisconsin-Madison, Madison, WI 53706, United States}
\author{Jon Ramos}
\affiliation[University of Wisconsin-Madison MS\&E]
{Department of Materials Science and Engineering, University of Wisconsin-Madison, Madison, WI 53706, United States}
\author{Sara Yarworth}
\affiliation[Water Science and Engineering Laboratory]
{Water Science and Engineering Laboratory, Department of Civil and Environmental Engineering, University of Wisconsin-Madison, Madison, WI 53706, United States}
\author{James Lazarcik}
\affiliation[Water Science and Engineering Laboratory]
{Water Science and Engineering Laboratory, Department of Civil and Environmental Engineering, University of Wisconsin-Madison, Madison, WI 53706, United States}
\author{Daniel A. Rhodes}
\email{darhodes@wisc.edu}
\affiliation[University of Wisconsin-Madison MS\&E]
{Department of Materials Science and Engineering, University of Wisconsin-Madison, Madison, WI 53706, United States}
\alsoaffiliation[UW-Madison Physics]
{Department of Physics, University of Wisconsin-Madison, Madison, WI 53706, United States}

\title[An \textsf{achemso} demo]
  {Deintercalation Equilibria Control the Composition, Structure, and Electronic Quality in the Synthesis of 2\textit{M}-WS\textsubscript{2}}

\begin{document}

\begin{tocentry}

\includegraphics[width=1\linewidth]{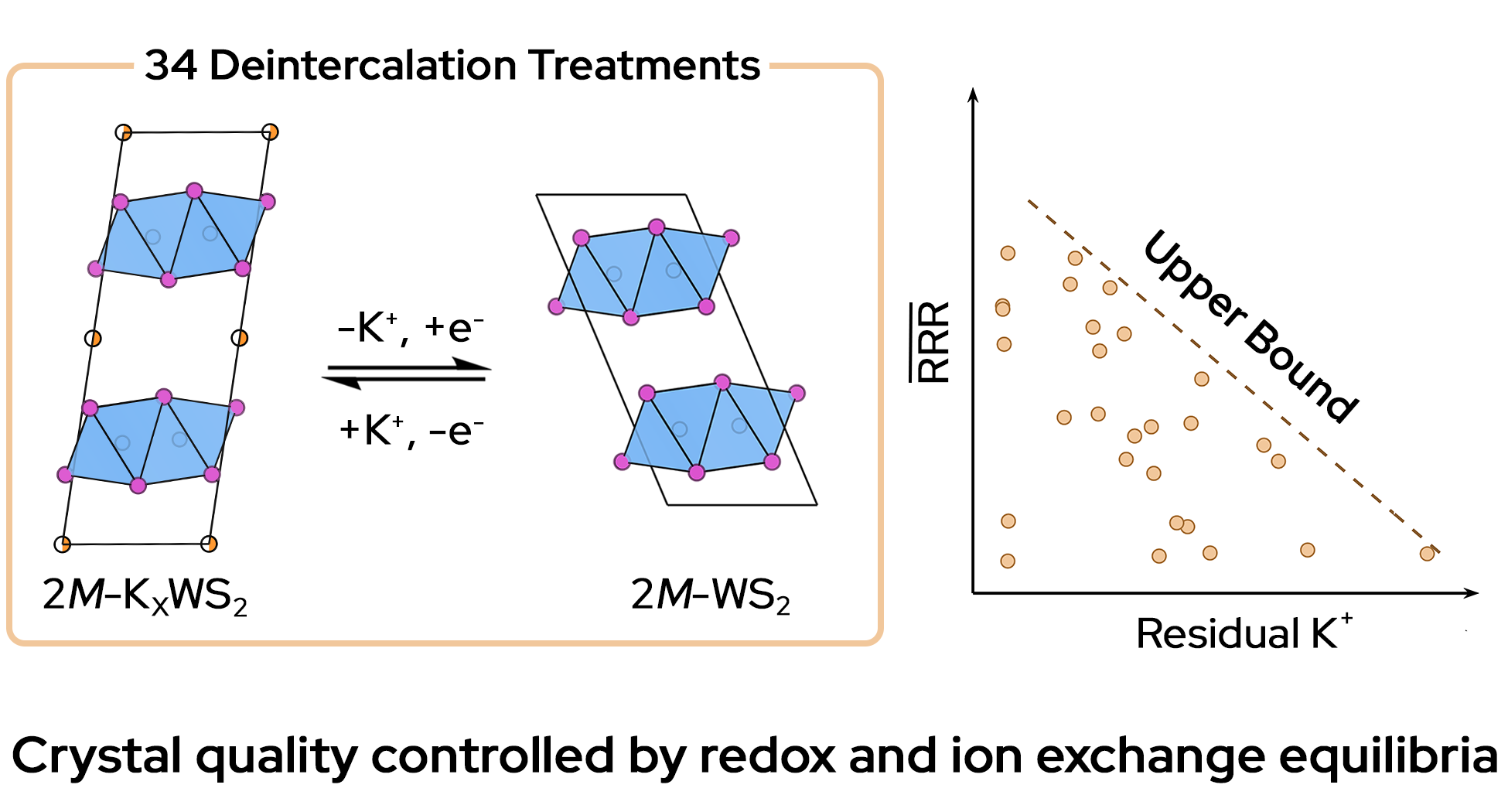}

\end{tocentry}

\begin{abstract}
Alkali-metal intercalation-deintercalation (AMID) is an established route to access metastable layered materials such as 2$M$-WS$_2$. Mechanistic study of AMID has focused on intercalation; how deintercalation proceeds and influences crystal properties is comparatively poorly understood. Here, we assess 34 deintercalation treatments of 2$M$-K$_x$WS$_2$ and find that K$^+$ removal appears to proceed through coupled redox and ion-exchange equilibria, whose positions set the composition, structure, and electronic quality of the products. Across a series of neat solvents, the extent of K$^+$ deintercalation increased with solvent polarity and was driven to near-completion upon addition of nonoxidizing crown ether chelators, regardless of solvent polarity. Spontaneous reintercalation occurred when 2$M$-WS$_2$ was placed in dilute aqueous solutions of the strong base KOH but not the neutral salt KCl. These results suggest that the extent of deintercalation is governed by the stability of K$^+$ in solution, rather than oxidant strength alone, consistent with a reaction under thermodynamic control. Treatments that bias the deintercalation equilibrium most strongly toward the products while avoiding deleterious side reactions produce the highest quality 2$M$-WS$_2$. A six-hour treatment with acidified K$_2$Cr$_2$O$_7$ gave the highest average residual resistivity ratio ($\overline{RRR}$). Residual K$^+$, quantified by ICP-MS, was found in all treated samples and appeared to impose an upper bound on $\overline{RRR}$. Treatments yielding crystals with $\overline{RRR}$ below this bound did not always afford 2$M$-WS$_2$; we observed two new structures in bleach-treated samples and signatures of alternative phases under other conditions. This work demonstrates that deintercalation is not merely a processing step in AMID but a mechanistically rich reaction that might be rationally tuned to improve and control metastable materials.
\end{abstract}

\newpage
\section{Introduction}
\begin{figure*}[tp]
   \centering
  \includegraphics[width=0.95\linewidth]{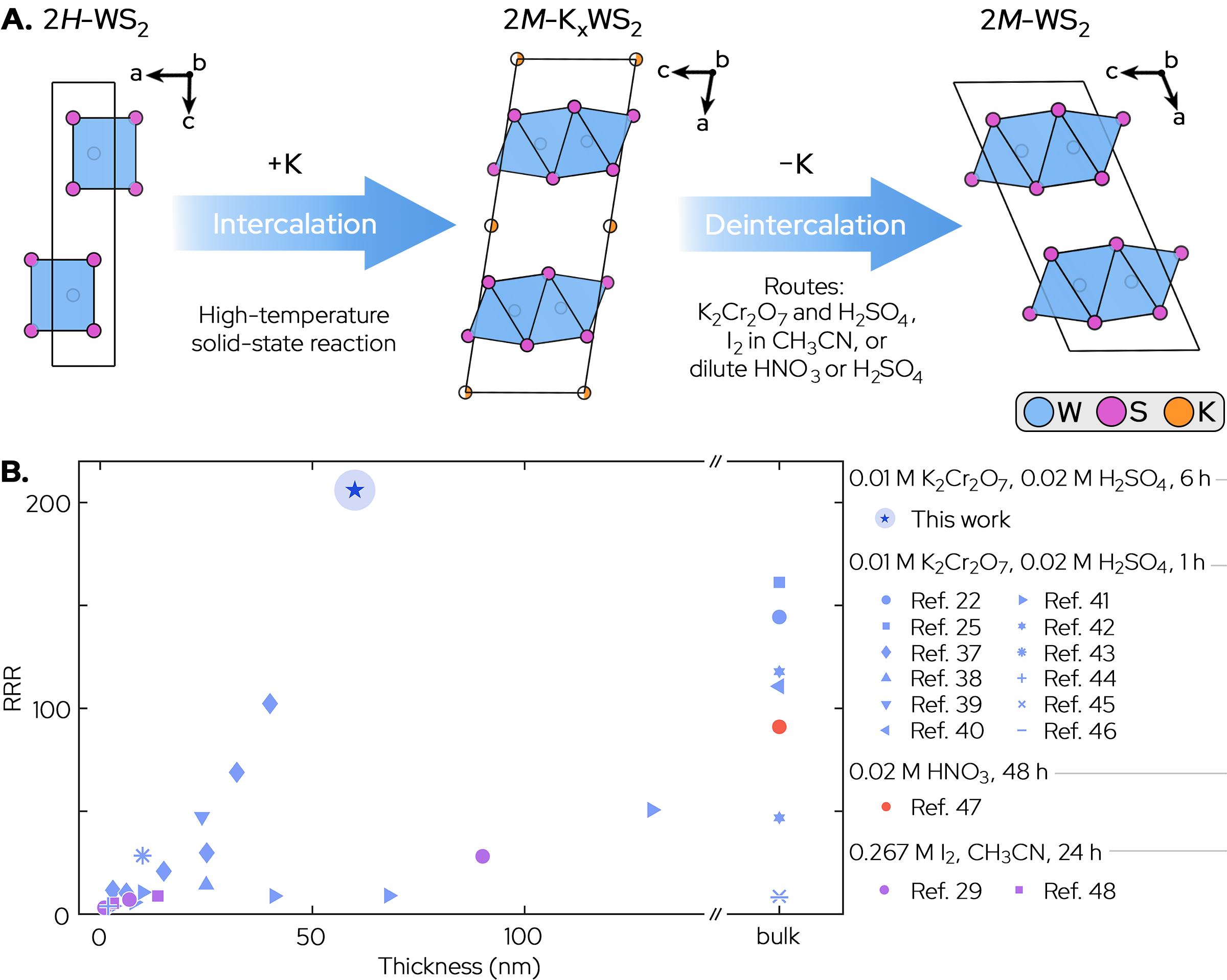}
   \caption{(A) Synthetic route for 2$M$-WS$_2$, showing the thermodynamically stable~\cite{VanArkel1926}, intermediate~\cite{Song2023b}, and final structures.~\cite{Fang2019} (B) Summary of RRRs for 2$M$-WS$_2$ reported in the literature.~\cite{Li2021,Fang2019,Samarawickrama2024,Samarawickrama2021,Che2019,Wang2020,Zhou2026,Yang2023,Mao2025,Gan2026,Zhao2021,Zhao2021b,Xie2026,Lai2021,Ji2022} RRRs were calculated based on published data to ensure a consistent method (details in Supporting Information).}
    \label{Fig1}
\end{figure*}

Intercalation and subsequent deintercalation of alkali metals are widely used to access metastable layered structures with useful or exotic properties~\cite{Takada2003,Zhou2016,Yu2018,Song2021,Yang2026}, usually at the cost of increased disorder in the product materials~\cite{Bhaskar2021,Roseler2025,Liers2025}. Disorder can degrade electronic properties such as carrier mobilities~\cite{Ma2014} and obscure signatures of sensitive quantum phenomena, such as the quantum spin Hall state~\cite{Shojaei2018,Lunczer2019}. Minimizing disorder is therefore required to unambiguously assess the intrinsic electronic behaviors of quantum materials. Although alkali-metal intercalation-deintercalation (AMID) is a well-established approach for obtaining metastable layered materials, mechanistic study of this technique has largely focused on understanding the intercalation reaction and its effect on the structures and properties of resulting materials.~\cite{Rudorff1959,Acrivos1979,Friend1987,Ohuchi1989,Muscher2021,Xu2024,Cassidy2025,Hossain2025} The deintercalation process and its impact on crystal quality remain comparatively underexplored. 
\\ 
\indent
One prominent material synthesized via AMID is the metastable transition metal dichalcogenide (TMD) 2$M$-WS$_2$. This material has the highest intrinsic superconducting transition temperature reported among TMDs ($T_\textrm{c}$ $\approx$ 8.8 K)~\cite{Li2024} and an in-plane upper critical field well above the Pauli limit, a violation attributed to spin–orbit–parity coupling.~\cite{Zhang2022} Moreover, signatures of topological superconductivity (TSC) have been reported in 2$M$-WS$_2$, including a superconducting gap developed on the topological surface states~\cite{Li2021} and a zero-bias conductance peak in vortex cores~\cite{Yuan2019}. Because the superconducting properties of 2$M$-WS$_2$ are sensitive to disorder~\cite{Ghimire2024}, the unambiguous identification of TSC-related phenomena is expected to require samples with minimal defects. 
\\ 
\indent
2$M$-WS$_2$ is commonly synthesized via a two-step AMID process (Figure \ref{Fig1}A). First, 2$H$-WS$_2$ is intercalated with an alkali metal, typically K, via a high-temperature solid-state reaction with K$^0$ or K$_2$S$_2$.~\cite{Fang2019,Guguchia2019} The resulting 2$M$-K$_x$WS$_2$ adopts a distorted octahedral W coordination rather than the trigonal prismatic W coordination of 2$H$-WS$_2$. Next, the alkali metal is deintercalated at room temperature, which preserves the distorted octahedral W coordination and yields metastable 2$M$-WS$_2$. Deintercalation has been reported to proceed via a redox event wherein [WS$_2$]$^{x-}$ moieties in 2$M$-K$_x$WS$_2$ are oxidized by an acid or other oxidizing agent, resulting in the ejection of K$^+$ from the crystal.~\cite{Song2023} The process has been suggested to proceed to a single end point, complete K$^+$ removal, at a rate set by the reduction potential of the oxidant.~\cite{Song2023} Three deintercalation methods for 2$M$-K$_x$WS$_2$ have been reported: submersion of samples in dilute acidified K$_2$Cr$_2$O$_7$ solution for one to two hours~\cite{Wypych1992,Fang2019}, I$_2$ in MeCN for 24 hours~\cite{Lai2021}, or dilute mineral acid (H$_2$SO$_4$ or HNO$_3$) for 48 hours~\cite{Song2023}. Other deintercalation methods have been mentioned for K$_x$WS$_2$, including electrochemical oxidation or use of FeCl$_3$ as an oxidizing agent, but the identity of the final product generated by these methods remains unclear~\cite{Schollhorn1984}. 
\par
Across other layered material systems, the deintercalation of alkali metals has been accomplished using a far wider array of treatments,~\cite{Ohtani1993,Wizansky1989,Takada2003,Liu2005,Xie2025} yet for 2$M$-WS$_2$ few conditions have been tested, and the effectiveness of fewer still have been directly compared. To our knowledge, only one such comparison has been reported, in which three deintercalation treatments were assessed.~\cite{Song2023} This study primarily evaluated treatments by the measured superconducting fraction of deintercalated samples and concluded that treatment of 2$M$-K$_{0.5}$WS$_2$ with dilute mineral acids alone yielded higher quality crystals than the traditional treatment with a strongly oxidizing acidified K$_2$Cr$_2$O$_7$ solution. What distinguishes effective from damaging treatments, and whether new conditions might yield higher-quality crystals remain poorly understood.
\par
%
\indent
In this work, we assessed the effectiveness of 34 deintercalation treatments of 2$M$-K$_x$WS$_2$ spanning strong oxidants, dilute acids, neat solvents, nonoxidizing chelating additives, or mixtures thereof in generating high quality 2$M$-WS$_2$. Substantial deintercalation occurred under almost all conditions tested, but was always incomplete. Samples treated with pure solvent exhibited K$^+$ removal, with increased solvent polarity corresponding to less K$^+$ remaining in the samples. Addition of nonoxidizing crown ether chelators to otherwise pure solvent led to nearly complete removal of K$^+$, even in nonpolar solvents that drove minimal deintercalation when neat. Deintercalated samples reintercalated spontaneously at room temperature upon exposure to solutions of the strong base KOH but not when immersed in a solution of the neutral salt KCl. Taken together, these results suggest that K$^+$ deintercalation is under thermodynamic control. Residual K$^+$, controlled by the equilibrium position of the deintercalation reaction, appears to be one factor governing the maximum observed average residual resistivity ratio ($\overline{RRR}$) of 2$M$-WS$_2$ and imposed an upper bound on $\overline{RRR}$. Several treatments that removed substantial K$^+$ nonetheless yielded crystals with $\overline{RRR}$ well below that of the best performing treatments with similar levels of K$^+$ removal, indicating that minimal residual K$^+$ is necessary but not sufficient for high crystal quality. We attribute this result to competing reactions that occur during the treatment and damage the lattice or potentially generate alternative phases. Among the conditions tested, treatment with 0.01 M K$_2$Cr$_2$O$_7$ in 0.02 M H$_2$SO$_4$ for 6 hours afforded crystals with the highest $\overline{RRR}$ (165). One sample prepared by this method had an RRR of 206, higher than any value previously reported for 2$M$-WS$_2$ to the best of our knowledge (Figure \ref{Fig1}B). 
\par
Our findings are inconsistent with the framing of deintercalation as a kinetically controlled process driven by [WS$_2$]$^{x-}$ oxidation alone. Instead, increased stabilization of the products relative to the reactants appears to promote more complete deintercalation of 2$M$-K$_x$WS$_2$ and higher crystal quality if deleterious side reactions are minimized. These results underscore that mechanistic understanding of the elementary steps underlying deintercalation could enable rational application of AMID to materials synthesis, granting greater control over resulting crystal quality and structure.  

\section{Results and discussion}
\subsection{Mechanistic Studies}
\begin{figure}[!htbp]
   \centering
  \includegraphics[width=\linewidth]{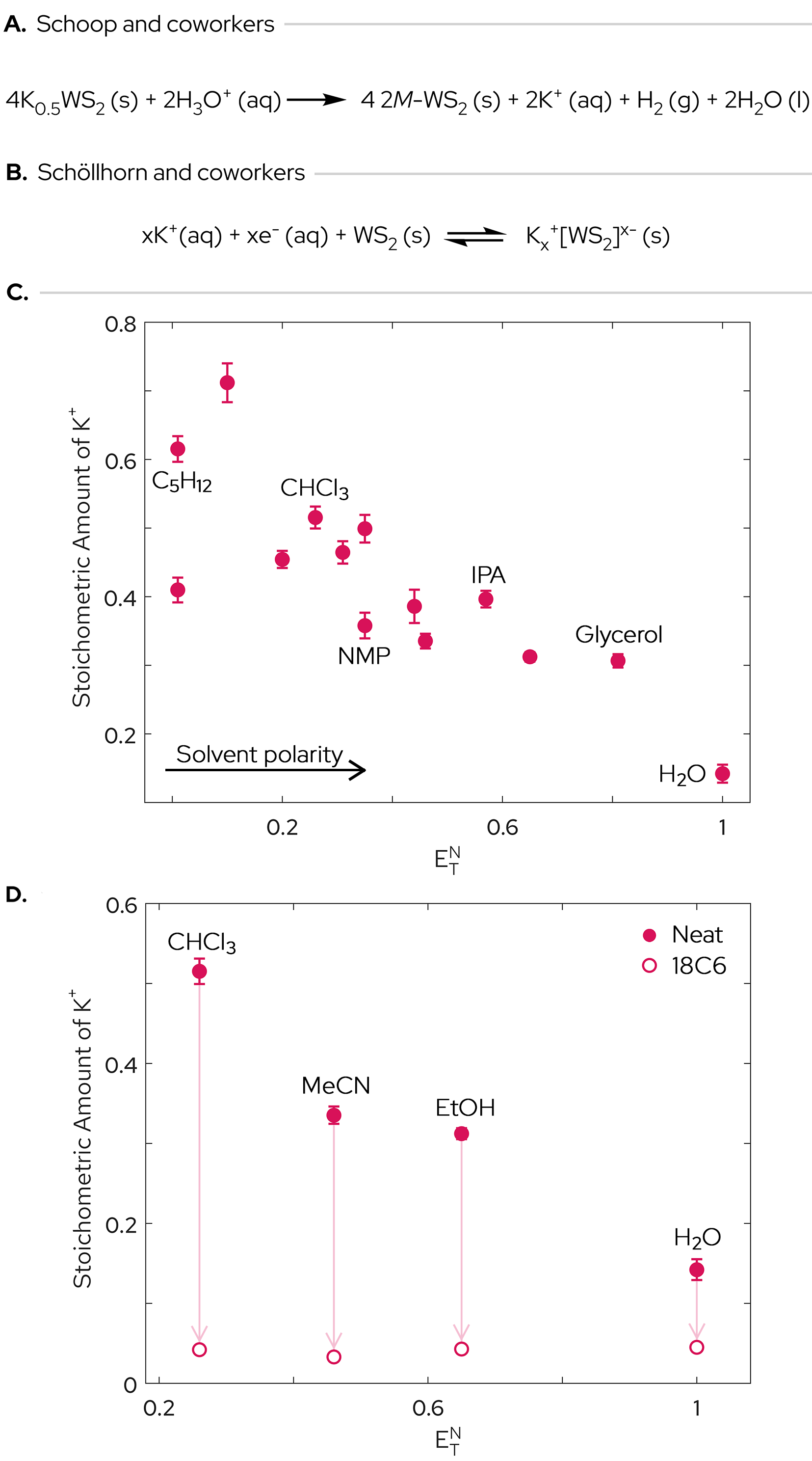}
   \caption{2$M$-K$_x$WS$_2$ deintercalation chemistry. (A) Proposed acid-mediated deintercalation equation.\cite{Song2023} (B) General AMID equilibrium half-reaction applied to K$_x$WS$_2$.\cite{Schollhorn1980} (C) K$^+$ remaining (EDS) after 48 hours vs E$_T^N$ without and (D) with 0.473 M 18C6. EDS values near the detection limit are less reliable; data in Table S1.}
    \label{Fig2}
\end{figure}

 In most reported 2$M$-WS$_2$ syntheses, the deintercalation reaction was either not described or was summarized simply as the oxidation of K$_x$WS$_2$.~\cite{Fang2019,Samarawickrama2024,Samarawickrama2021,Li2021,Che2019,Wang2020,Zhou2026,Yang2023,Zhao2021,Zhao2021b} Deeper chemical insight into the deintercalation was provided by Schoop and coworkers, who describe the process as the oxidation of [WS$_2$]$^{x-}$ with concomitant reduction of H$_3$O$^+$ to H$_2$ gas and ejection of K$^+$ into the reaction solution (Figure \ref{Fig2}A).~\cite{Song2023} In this mechanistic framework, deintercalation is a process that results in quantitative K$^+$ removal provided sufficient time. The necessary time for complete deintercalation depends on the reduction potential of the oxidizing agent involved in the reaction. This work established a redox couple for deintercalation but did not explicitly identify whether the process was kinetically or thermodynamically controlled. Differentiation between the two types of reaction control based on the products of the acid-assisted deintercalation may be challenging as the evolution of H$_2$ gas would drive the reaction towards complete deintercalation given enough time even if the transformation were an equilibrium, thus mimicking irreversibility.
 
\par
The mechanism of AMID reactions across a range of TMDs was examined by Sch\"{o}llhorn and coworkers.~\cite{Schollhorn1980} They found that, upon alkali-metal introduction, TMD layers were reduced, becoming negatively charged, and cations intercalated into the interlayer gallery to balance the charge (Figure \ref{Fig2}B).~\cite{Schollhorn1974,Schollhorn1975} Intercalated phases generated in this way function as polyelectrolytes. Placement of alkali-metal-intercalated TMDs into oxidative solutions resulted in deintercalation of cations from the interlayer gallery, yielding neutral TMD layers. In the solid-state synthesis of 2$M$-K$_x$WS$_2$, K$^0$ is presumably oxidized to K$^+$ through transfer of electrons to WS$_2$ moieties, thereby lowering the average formal oxidation state of W from +4 to +4-$x$ and resulting in K$^+$ intercalation. Cyclic voltammetry indicated that intercalation of hydrated K$^+$ into 2$M$-WS$_2$ proceeded through four intermediate phases, each with a corresponding oxidation peak on the reverse sweep.\cite{Wypych1992} In situ neutron diffraction on similar AMID TMD phases suggested that oxidation and reduction proceeded through the same staged intermediate structures.\cite{Schollhorn1980,Riekel1980} This work established chemical reversibility for these systems, but whether the intercalation-deintercalation reaction is controlled by thermodynamics or kinetics remains an open question.
\par 
If deintercalation of 2$M$-K$_x$WS$_2$ is governed by thermodynamics, the residual K$^+$ content in the lattice should depend on how strongly the treatment solution stabilizes K$^+$ or can drive the equilibrium forward. To examine this possibility, we evaluated the effectiveness of neat solvents of differing polarity for removing interstitial K$^+$. K$^+$ remaining after treatment was measured by energy-dispersive spectroscopy (EDS). When 2$M$-K$_x$WS$_2$ was placed in nanopure water for 48 hours, considerable K$^+$ was removed, suggesting that oxidation of [WS$_2$]$^{x-}$ is facile and can likely be driven by water or dissolved oxygen. If deintercalation is reversible, changing the solvent to one less capable of stabilizing K$^+$ should decrease the equilibrium constant, resulting in reduced K$^+$ deintercalation. Polar solvents, such as H$_2$O, can engage in strong ion-dipole interactions that stabilize cations, whereas nonpolar solvents, like C$_5$H$_{12}$, offer little stabilization to cations and disfavor charge separation. We tested the ability of 14 solvents of diverse polarity, held at room temperature, as deintercalation treatments for 2$M$-K$_x$WS$_2$. Solvent polarity was quantified using Reichardt's dye parameter (E$_T$(30)),\bibnote{E$_T$(30) is defined as the molar electronic transition energy (in kcal/mol) of the negatively solvatochromic betaine dye 30, calculated from its longest-wavelength charge-transfer absorption.} normalized on a scale from tetramethylsilane to water (E$_T^N$).~\cite{Reichardt1994} High E$_T^N$ values correspond to high solvent polarity. We found that increased solvent polarity correlated with reduced residual K$^+$ (Figure \ref{Fig2}C). The highly nonpolar solvent C$_5$H$_{12}$ (E$_T^N$ = 0.01) removed minimal K$^+$, while the much more polar glycerol (E$_T^N$ = 0.81) removed substantially more K$^+$. This trend is consistent with deintercalation being a thermodynamically-controlled process--conditions that better stabilize K$^+$ outside the interlayer gallery favor its removal from the crystal.  
\par
To further probe the role of K$^+$ stability in the reaction outcome, we treated 2$M$-K$_x$WS$_2$ with nonoxidizing, chemically-inert crown ether chelators dissolved in solvents of high (H$_2$O), moderate (EtOH and MeCN), and low (CHCl$_3$) polarity. Crown ethers such as 18-crown-6 (18C6) are multidentate cyclic oligoethers known to strongly stabilize alkali-metal cations through noncovalent ion–dipole interactions~\cite{Poonia1979}. Complexation of K$^+$ by 18C6 should lower the free energy of the cation in solution, rendering transfer of intercalated K$^+$ out of the crystal more thermodynamically favorable. When 2$M$-K$_x$WS$_2$ was placed in a concentrated aqueous solution of 18C6 for 48 hours, substantially more K$^+$ was removed relative to treatment with neat water (Figure \ref{Fig2}D). In nonpolar solvents that removed little K$^+$ when neat, addition of 18C6 resulted in significantly increased deintercalation. As crown ethers are electron-rich and very poor oxidants, the marked increase in K$^+$ removal upon their addition to the treatment solutions suggests that the enhanced deintercalation results from greater stabilization, as opposed to faster oxidation.
\par
\begin{figure}[!htbp]
   \centering
  \includegraphics[width=\linewidth]{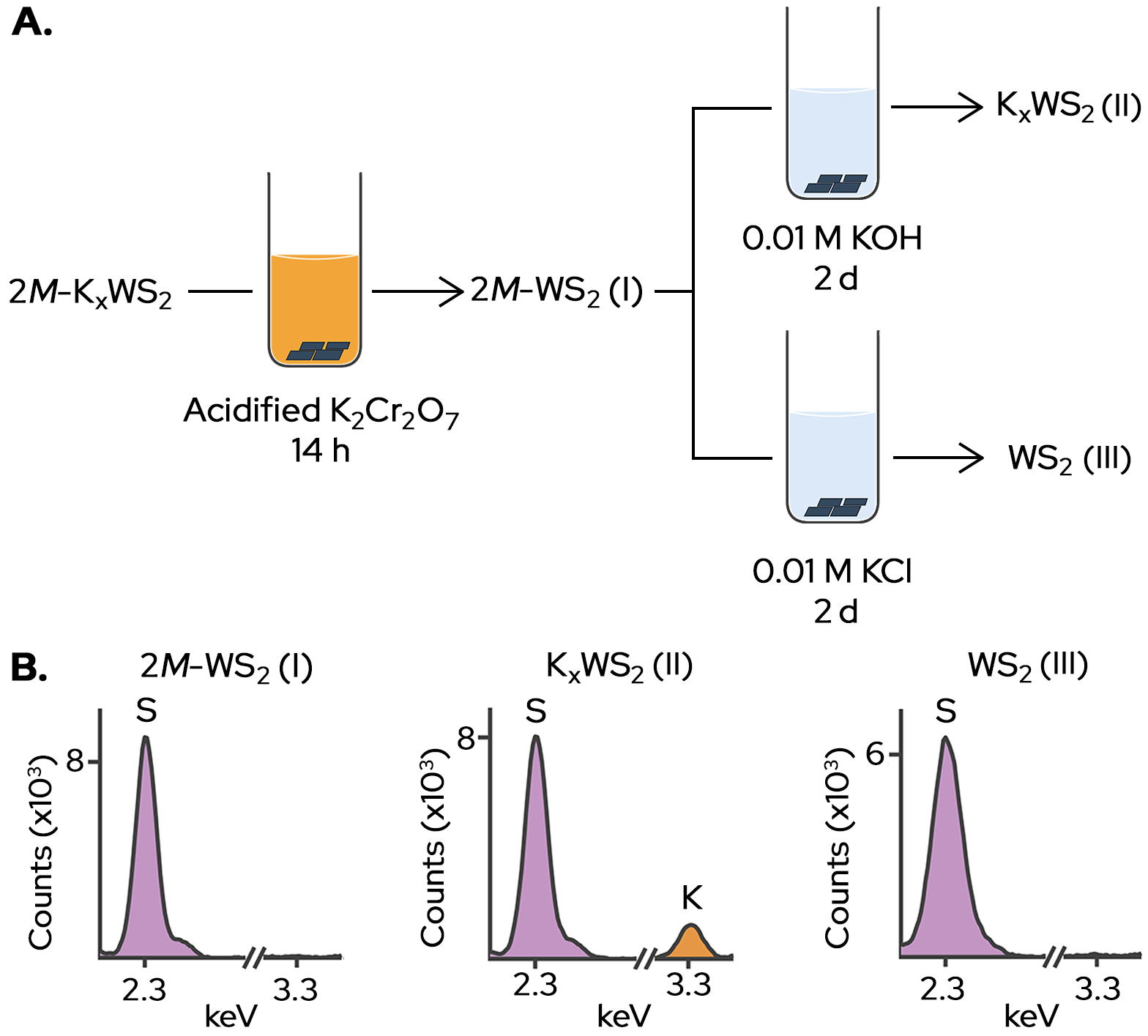}
   \caption{Deintercalation and reintercalation experiments. (A) Summary of treatments and (B) magnified view of the S K$\alpha$ and K K$\alpha$ peaks from EDS spectra for each sample. For full spectra, see Figures S1--S3.}
    \label{Fig3new}
\end{figure}

A thermodynamically controlled deintercalation should be reversible, with K$^+$ spontaneously reintercalating into 2$M$-WS$_2$ under conditions where K$^+$ are less stable in solution relative to within the interlayer gallery. To assess this hypothesis, we exposed deintercalated 2$M$-WS$_2$ samples (prepared via exposure to 0.01 M K$_2$Cr$_2$O$_7$ in 0.02 M H$_2$SO$_4$ for 14 hours (Figure S1)) to dilute (0.01 M) aqueous KOH or KCl solutions (Figure \ref{Fig3new}A). KOH is a strong base and unstable (pKa of the conjugate acid, H$_2$O = 14.0), whereas KCl is an extremely weak, highly stable base (pKa of conjugate acid, HCl = $-7.3$).~\cite{williams_pka_2022} Exposure of the deintercalated samples to the KOH solution resulted in reintercalation, yielding K$_x$WS$_2$ after 48 hours of treatment (Figure \ref{Fig3new}B). Exposure of deintercalated 2$M$-WS$_2$ to dilute KCl resulted in minimal reintercalation. Treatment of deintercalated samples with saturated aqueous solutions of KOH or KCl resulted in similar outcomes: concentrated KOH-treated samples were reintercalated whereas KCl-treated samples largely remained deintercalated (Figures S4 and S5). These results are consistent with the hypothesis that the stability of the products and reactants control the outcome of the reaction. These data and the results of the solvent polarity and chelator experiments strongly suggest that the deintercalation reaction is under thermodynamic control. 

\subsection{Defining Crystal Quality}
\begin{figure}[!htbp]
   \centering
  \includegraphics[width=\linewidth]{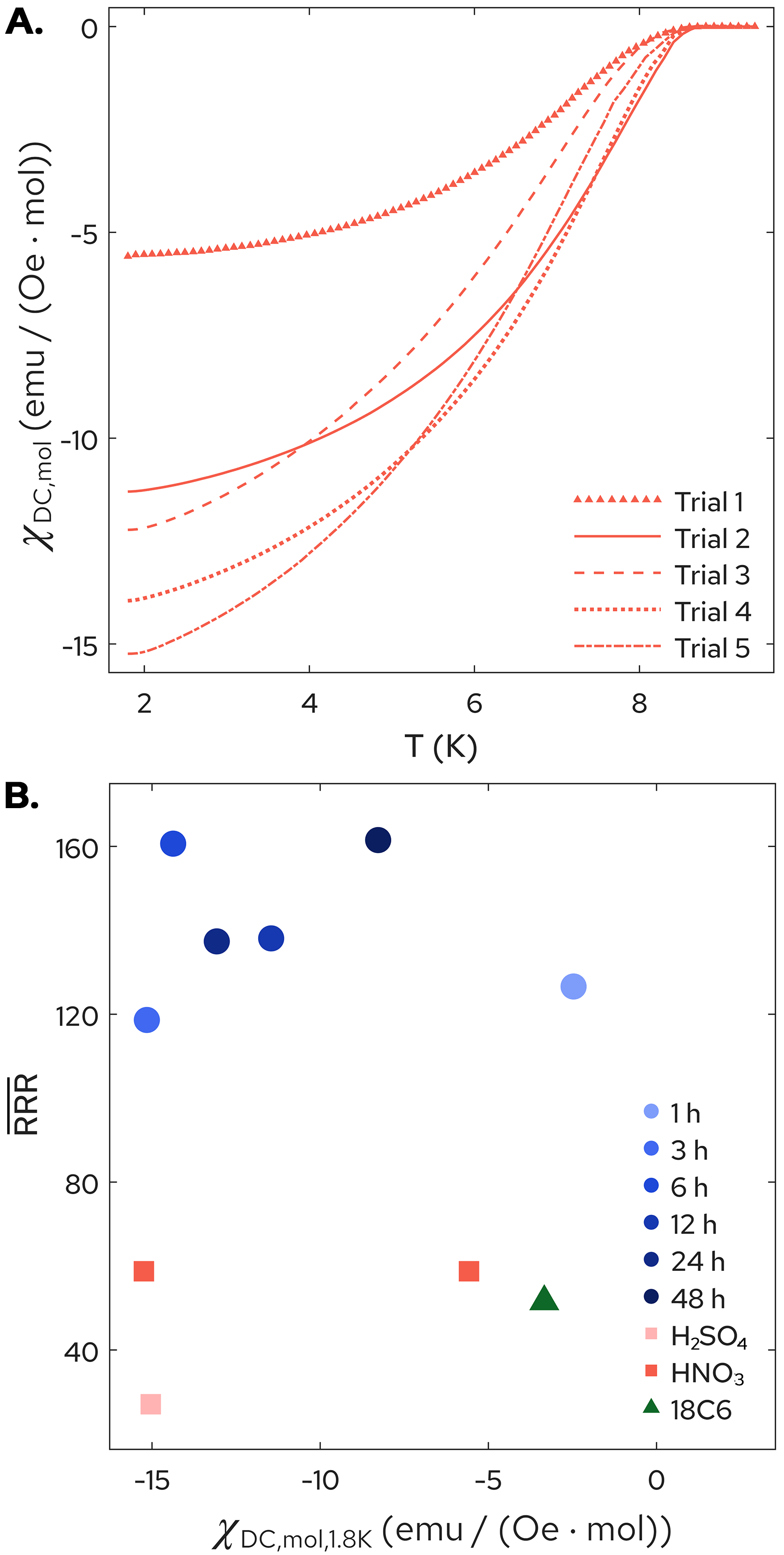}
   \caption{Evaluation of crystal quality metrics. (A) $\chi$ vs T for five different samples of 2$M$-WS$_2$ treated with 0.02 M HNO$_3$ for 48 hours. (B) Correlation between $\overline{RRR}$ and $\chi_{1.8 \mathrm{K}}$ for select conditions. $\chi$ vs T curves are included as Figures S6 and S7. Unless noted, all treatments are aqueous, 48 hours, or use 0.01 M K$_2$Cr$_2$O$_7$ in 0.02 M H$_2$SO$_4$; other conditions are 0.01 M H$_2$SO$_4$, 0.02 M HNO$_3$, or 0.473 M 18C6.}
    \label{Fig3}
\end{figure}

 Two metrics have been utilized to evaluate the crystal quality of 2$M$-WS$_2$: RRR and superconducting volume fraction~\cite{Fang2019,Song2023}. RRR is calculated by dividing the resistivity ($\rho$) at room temperature ($\rho_{300 K}$) by the resistivity just before the superconducting transition ($\rho_{T_\textrm{c}}$) for a sample (see Supporting Information). A higher RRR indicates a reduced defect density, as scattering from defects or impurities increases $\rho_{T_\textrm{c}}$. Superconducting volume fraction is assessed by zero-field cooled magnetic susceptibility ($\chi$). A more negative $\chi_{1.8 \mathrm{K}}$ corresponds to a stronger Meissner response per unit volume, indicating a higher superconducting fraction and potentially greater phase purity.  
\\ 
\indent
We observed $\chi_{1.8 \mathrm{K}}$ to exhibit high sample-to-sample variation (Figure \ref{Fig3}A). Across five different samples of 2$M$-WS$_2$ deintercalated with 0.02 M HNO$_3$ for 48 hours, we observed a range for $\chi_{1.8 \mathrm{K}}$ of 9.7 $\mathrm{emu}\,/(\mathrm{Oe}\cdot\mathrm{mol})$, representing over 75\% of the 12.8 $\mathrm{emu}\,/(\mathrm{Oe}\cdot\mathrm{mol})$ range for all samples tested. As each collection utilized between 8.5 to 25.5 mg of sample, a high degree of averaging should be present, making the extent of variability more surprising. Similar variation appeared in a previous study of 2$M$-WS$_2$,~\cite{Song2023} that used $\chi_{1.8 \mathrm{K}}$ as a primary measure of crystal quality and found a difference of around 14.8 $\mathrm{emu}\,/(\mathrm{Oe}\cdot\mathrm{mol})$ ($\approx$47\% difference) between the two 0.02 M HNO$_3$ trials reported. Such high variability adds uncertainty to direct comparisons of crystal quality using this method.
\par
Although RRR and $\chi_{1.8 \mathrm{K}}$ have both been used as metrics for crystal quality, we found little correlation between the two (Figure \ref{Fig3}B). Samples with comparable $\chi_{1.8 \mathrm{K}}$ differed in $\overline{RRR}$ by nearly an order of magnitude, and samples with commensurate $\overline{RRR}$ differed in $\chi_{1.8 \mathrm{K}}$ by a factor of three. This may be explained by each technique being sensitive to different, though overlapping, aspects of crystal quality. For example, a material could contain a significant non-superconducting phase alongside a very low-defect superconducting phase. $\chi_{1.8 \mathrm{K}}$ may reflect phase purity, whereas RRR captures the electronic behavior as affected by defects in the crystal lattice. As 2$M$-WS$_2$ is of interest as a TSC, we focus on electronic quality and use $\overline{RRR}$ to quantify crystal quality throughout this work.
\subsection{Impact of Deintercalation Reaction on Electronic Properties}
\begin{figure}[!htbp]
   \centering
  \includegraphics[width=\linewidth]{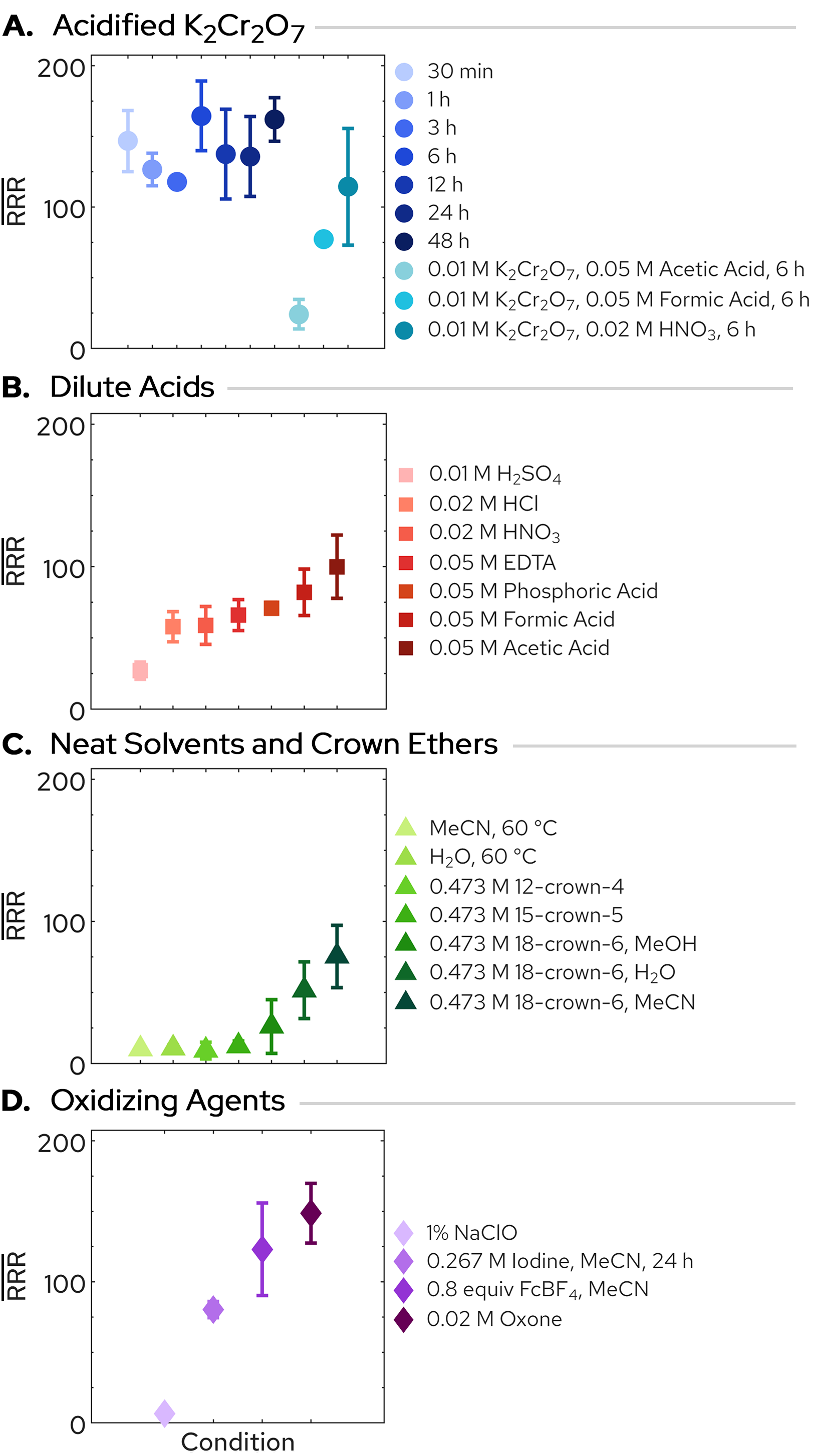}
   \caption{$\overline{RRR}$ for (A) acidified K$_2$Cr$_2$O$_7$, (B) dilute acids, (C) neat solvents and crown ether solutions, and (D) oxidizing treatments. Unless otherwise noted, the treatment time used was 48 hours and K$_2$Cr$_2$O$_7$ solutions were 0.01 M K$_2$Cr$_2$O$_7$ in 0.02 M H$_2$SO$_4$.}
    \label{Fig4}
\end{figure}

To evaluate the effect of the deintercalation route on crystal quality, we measured $\overline{RRR}$ for 2$M$-WS$_2$ crystals obtained using 34 different deintercalation treatments. 2$M$-WS$_2$ crystals were exfoliated with tape and transferred onto Pt electrodes on SiO$_2$ (see Supporting Information for details). Significant variation in RRR was observed from sample to sample within a single treatment (Figure S8). RRR is known to vary from crystal to crystal within a single growth and between samples with different thicknesses. To minimize variability due to sample thickness, all samples measured were approximately 50 to 80 nm thick. By calculating an $\overline{RRR}$ across multiple samples per treatment condition, we observe statistically significant differences between deintercalation conditions. Our results are grouped into four categories: acidified K$_2$Cr$_2$O$_7$ solutions, dilute acids, crown ethers and neat solvents, and alternative oxidants (Figure \ref{Fig4}). 
\\ 
\indent
Acidified K$_2$Cr$_2$O$_7$ treatments were evaluated across exposure durations from 30 minutes to 48 hours. 2$M$-WS$_2$ is typically prepared through exposure to K$_2$Cr$_2$O$_7$ in H$_2$SO$_4$ for one hour.~\cite{Fang2019,Samarawickrama2024} Extending the standard one-hour treatment to six hours yielded the highest $\overline{RRR}$ among all conditions tested (Figure \ref{Fig4}A), showing that modest optimization of common methods could substantially improve crystal quality. Crystals treated for longer than six hours had a comparable $\overline{RRR}$ but more frequently featured a depressed $T_\textrm{c}$ (Figures S9 and S10). The depression in $T_\textrm{c}$ may arise from progressive degradation of the crystal during exposure to the treatment solution. If such damage occurs, the shorter-exposure samples are presumably also affected, but to a lesser degree. The six-hour treatment appears to represent the best compromise between sufficient deintercalation and this exposure-induced disorder. 
\\ 
\indent
To assess the prior claim that dilute acids produce higher-quality crystals than K$_2$Cr$_2$O$_7$ in H$_2$SO$_4$,~\cite{Song2023} we tested a range of dilute acids spanning from strong inorganic to weak organic acids. Surprisingly, all dilute acid treatments yielded crystals with a lower $\overline{RRR}$ than those prepared with K$_2$Cr$_2$O$_7$ in H$_2$SO$_4$ (Figure \ref{Fig4}B). The suppressed $\overline{RRR}$ may result from introduction of S vacancies and lattice disorder through competing side reactions with the acids.~\cite{Schollhorn1975} Although no H$_2$S or SO$_x$ evolution was reported during oxidation of 2$M$-K$_x$WS$_2$ in acid,~\cite{Song2023} the absence of gaseous products does not preclude lattice sulfur oxidation in TMDs.~\cite{Schippers1999} We found that changing acid identity or H$_3$O$^+$ concentration generally had limited impact on $\overline{RRR}$. A notable outlier was H$_2$SO$_4$; crystals from this treatment exhibited the lowest $\overline{RRR}$ of the acids tested and were visibly damaged (Figure S11). Similar damage was observed in treatments with oxalic acid, indicating that both strong and weak acids are capable of degrading samples (Figure S12). With respect to $\overline{RRR}$, we observed no advantage to preparing 2$M$-WS$_2$ under dilute acid conditions relative to stronger oxidizing conditions, suggesting that a treatment's capacity to drive deintercalation, rather than its oxidative mildness, may determine crystal quality. 
\par
That dilute H$_2$SO$_4$ alone yielded the poorest-quality crystals among the acids tested suggests that pairing K$_2$Cr$_2$O$_7$ with a different acid could afford higher-quality crystals. Under acidic conditions, K$_2$Cr$_2$O$_7$ dissociates to \ce{Cr2O7^2-} and hydrolyzes to form H$_2$CrO$_4$ in equilibrium with \ce{HCrO4^-} and \ce{CrO4^2-}~\cite{Sherrill1907}. We found that when K$_2$Cr$_2$O$_7$ was paired with HNO$_3$, formic, or acetic acid (p$K_a$ = -1.3, 3.8, 4.8, respectively), instead of H$_2$SO$_4$ (p$K_a$ = -3.0), the resulting 2$M$-WS$_2$ crystals were of lower quality.~\cite{williams_pka_2022} We surmise that these weaker acids generate insufficient chromic acid (p$K_a$ = -1.0) to drive deintercalation effectively. Across this acid series, crystal quality decreased with increasing p$K_a$. This trend is consistent with the equilibrium framework, as increasing the concentration of the strong oxidant displaces the reaction toward the deintercalated product, yielding higher-quality crystals.
\\
\indent
To probe milder conditions and their effect on $\overline{RRR}$, we examined crown ether solutions or heated neat solvents. $\overline{RRR}$ for all of these treatments remained below that of more strongly oxidizing conditions (Figure \ref{Fig4}C). Crystals treated with 60 $^{\circ}$C water or MeCN exhibited an $\overline{RRR}$ among the lowest values measured. For treatments with crown ethers, $\overline{RRR}$ increased with cavity size and was highest for 18C6, the crown ether best size-matched to K$^+$, consistent with K$^+$ stabilization promoting deintercalation. Within the 18C6 series, $\overline{RRR}$ varied with the solvent, but the variation tracked neither solvent polarity~\cite{Reichardt1994} nor the established solvent dependence of [K(18C6)]$^+$ complex stability.~\cite{Solovev1996} Combining 18C6 with the organic oxidizing agent tetracyanoethylene (in MeCN), or with acidified K$_2$Cr$_2$O$_7$, failed to produce higher-quality crystals (Figure S13). Reducing the concentration of aqueous 18C6 from 0.473 to 0.002 M further lowered crystal quality (Figure S13). These results suggest that while mild conditions can facilitate oxidation of 2$M$-K$_x$WS$_2$ and removal of K$^+$, too much K$^+$ may remain in the lattice at equilibrium to approach the crystal quality obtained with harsher treatments.
\\ 
\indent
To assess whether oxidant identity affects crystal quality, we screened oxidizing agents spanning a range of strengths and pH (Figure \ref{Fig4}D). Among the alternative oxidants tested, Oxone (\ce{KHSO5}$\cdot$\allowbreak\ce{0.5KHSO4}$\cdot$\allowbreak\ce{0.5K2SO4}) appeared to be the most promising, though its $\overline{RRR}$ was slightly lower than that achieved with the six-hour K$_2$Cr$_2$O$_7$ in H$_2$SO$_4$ treatment. A high $\overline{RRR}$ therefore does not appear to be specific to acidified K$_2$Cr$_2$O$_7$ treatment. We additionally found that K$^+$ deintercalation methods used for other materials, such as ferrocenium tetrafluoroborate (FcBF$_4$) in MeCN~\cite{Xie2025} or heated aqueous AlCl$_3$~\cite{Ohtani1993}, are effective at deintercalating 2$M$-WS$_2$ (Table S2), but provided mixed results for $\overline{RRR}$ (Figure S14). Alkali-metal deintercalation strategies appear transferable between layered materials, but their effect on crystal quality may not be, as the chemistry governing potential side reactions could be lattice-specific.
\par
The sensitivity of $\overline{RRR}$ to treatment duration and to alkali-metal intercalant identity was also assessed. Three conditions (1\% bleach, 0.02 M Oxone, and 0.01 M H$_2$SO$_4$) removed K$^+$ rapidly ($<$1 h) (Figure S15). For these conditions, reducing the treatment duration from 48 to 1 hour produced no significant change in $\overline{RRR}$ (Figure S13). Under these treatments the system appears to reach equilibrium rapidly, after which the sample is neither further deintercalated nor damaged, unlike the acidified K$_2$Cr$_2$O$_7$ solution in which extended exposure could depress $T_\textrm{c}$. Samples made from 2$M$-Rb$_x$WS$_2$~\cite{Hoff2024} rather than 2$M$-K$_x$WS$_2$ had a significantly lower $\overline{RRR}$ when prepared with K$_2$Cr$_2$O$_7$ in H$_2$SO$_4$ deintercalation treatments (Figure S13). This result suggests that intercalant identity could induce enduring, potentially deleterious changes to the lattice, or alter the governing deintercalation chemistry, indicating that both steps in the AMID process influence resultant crystal quality. 
\\ 
\indent
 The treatments tested here were selected to span distinct chemical environments, and together they suggest that crystal quality is controlled, at least in part, by a deintercalation equilibrium rather than by the oxidative mildness of the treatment. Conditions expected to displace the equilibrium farthest toward the deintercalated product tended to yield the highest $\overline{RRR}$, whereas the mildest oxidative treatments, which drive oxidation but may leave the equilibrium shifted further toward the intercalated starting material, underperformed. Several conditions that should strongly favor the deintercalated products yielded low $\overline{RRR}$ crystals, reflecting the possible presence of deleterious side reactions. Crystal quality thus appears to be governed by the bias of the equilibrium and the presence of side reactions that degrade 2$M$-WS$_2$.
 
\subsection{Impact of Deintercalation Treatment on Structure}
\begin{figure}[!htbp]
   \centering
  \includegraphics[width=\linewidth]{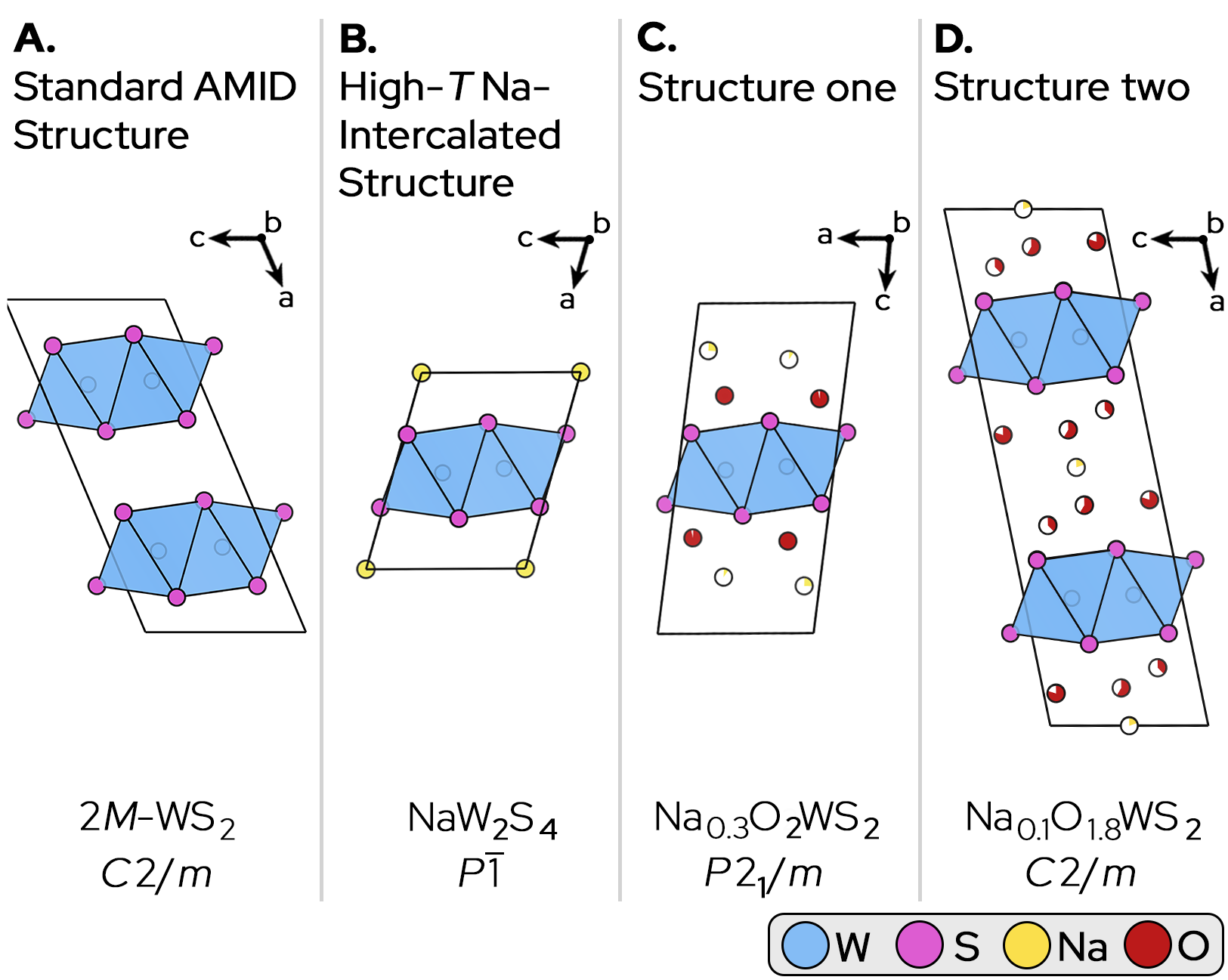}
   \caption{Structures observed in samples treated with 1\% bleach. (A) Reported AMID 2$M$-WS$_2$ structure~\cite{Fang2019}, (B) reported Na-intercalated WS$_2$ structure obtained with high-temperature solid-state synthesis,~\cite{Hoff2024} (C) structure one, and (D) structure two.}
    \label{structure}
\end{figure}

To identify the origin of the observed differences in $\overline{RRR}$, we examined the structure of deintercalated crystals produced with different treatments. We found that several treatments yielded crystals wherein the majority phase was not 2$M$-WS$_2$. Two new crystal structures were identified by single-crystal X-ray diffraction (SCXRD) from samples treated with 1\% bleach for 48 hours (Figure \ref{structure}). These alternative phases retain the distorted WS$_2$ layers observed in 2$M$-WS$_2$, but have expanded interlayer galleries and atomic arrangements consistent with reported hydrated alkali-metal TMD phases.~\cite{Schollhorn1974b,Lerf1977,Bos-Alberink1979,Mao2019,Liu2023,Zhang2024} Such hydrated compounds are described by the polyelectrolyte model, wherein the uptake of alkali metal cations is commensurate with co-intercalation of solvent, determined by the hydration energy and geometry of the cation.~\cite{Schollhorn1973,Schollhorn1974b,Schollhorn1975,Lerf1977} In structures one and two, the WS$_2$ layers are bounded above and below by planes of O atoms. A particularly short S--O bond distance (1.58 \AA) in structure one suggests covalent bonding between S and O. The O sites are presumed to be H$_2$O or OH$^-$, though any H present could not be located from these data, as its X-ray scattering contribution would fall below the noise level of our collection. The region enclosed by the O layers accommodates either one or two layers of partially occupied Na$^+$. The Na$^+$ assignment is supported by X-ray photoelectron spectroscopy (XPS) and EDS (Figures S16--S21). To the best of our knowledge, multiple distinct planes of alkali-metal intercalants have not previously been structurally resolved in TMDs, hydrated or anhydrous, although such arrangements are predicted to be feasible~\cite{Kuhne2018,Chepkasov2020,Chepkasov2022}, have been observed in other layered materials,~\cite{Lin2024,Pandey2025,Williams2026} and have a transition-metal analogue in TMDs~\cite{Roseler2026}. Spontaneous incorporation of guest species from room-temperature solutions is consistent with a cation-exchange reaction as described by the polyelectrolyte model.~\cite{Schollhorn1974b,Schollhorn1975,Lerf1977} As charge neutrality must be maintained, such reactions would primarily affect [WS$_2$]$^{x-}$  and contribute most strongly when the equilibrium position lies closer to the reactants. Further discussion of the structures, uncertainties in SCXRD solutions, and experimental details are included in the Supporting Information.
\begin{figure*}[tp]
   \centering
  \includegraphics[width=\linewidth]{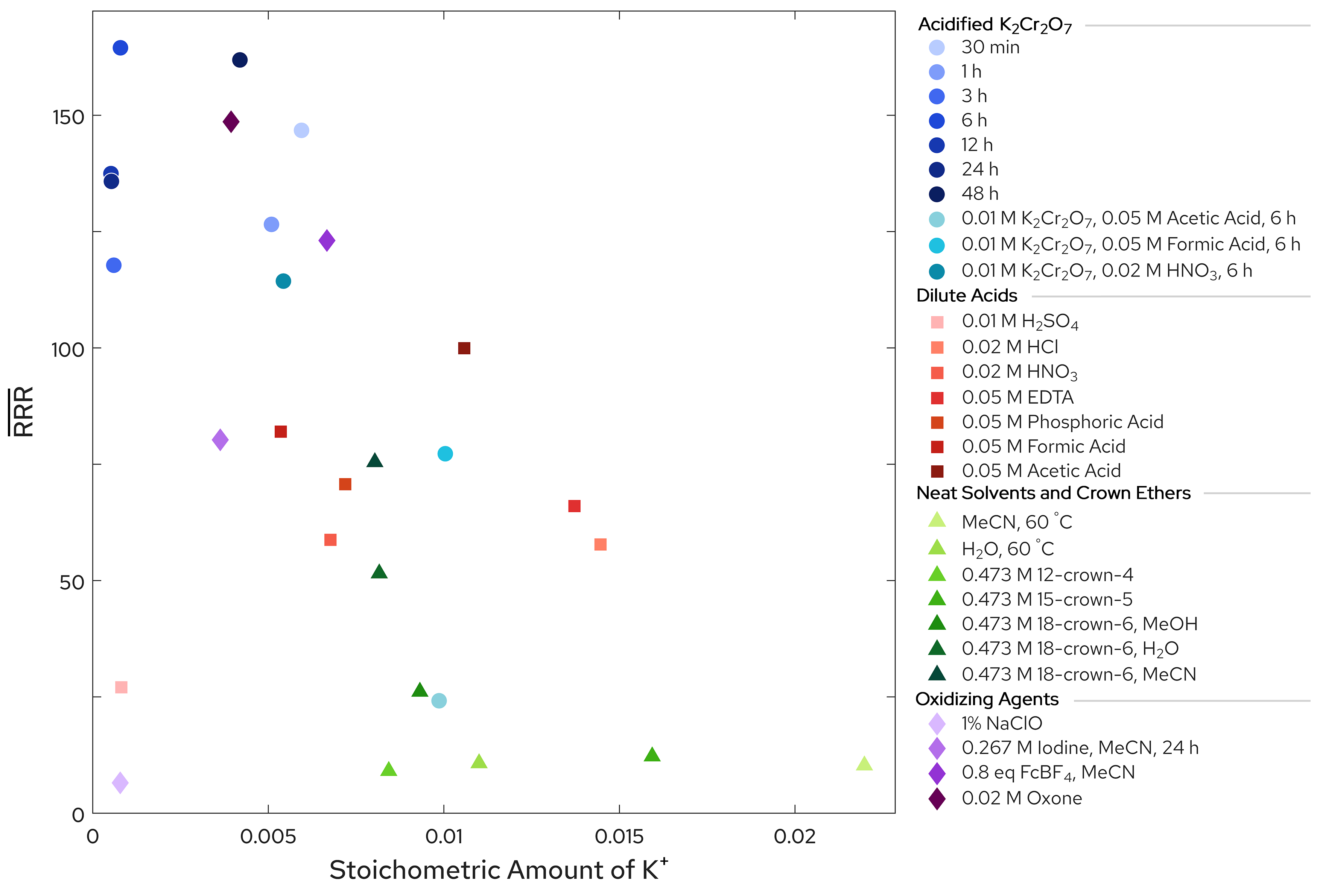}
   \caption{Correlation between residual K$^+$ (as measured by ICP-MS) and $\overline{RRR}$.}
    \label{ICP}
\end{figure*}
\par
Several treatments produced samples whose powder X-ray diffraction (PXRD) patterns and Raman spectra matched those of 2$M$-K$_x$WS$_2$ (Figures S22 and S23), despite elemental analysis indicating that K$^+$ had been largely removed from the samples (Table S2). In the absence of K$^+$, 2$M$-K$_x$WS$_2$ cannot account for these signatures, indicating the formation of alternative crystal structures for these samples. The bleach-treated samples described above are one such case. As these phases are nearly indistinguishable from 2$M$-K$_x$WS$_2$ by PXRD and Raman spectroscopy, such structures could be overlooked~\cite{Bierman2025}. Additionally, PXRD patterns of samples treated with bleach or heated water indicate that 2$M$-WS$_2$ and alternative structures can coexist within a single crystal (Figure S23). The phase purity therefore appears to vary with deintercalation treatment. Shifting the equilibrium toward the products could improve crystal quality and phase purity by minimizing the portion of remaining [WS$_2$]$^{x-}$ and limiting one route to alternative structure formation. 
\\ 
\indent
Crystals produced by some of the treatments displaying signs of alternative structure formation can exhibit suppressed superconductivity, including reduced $T_\textrm{c}$, partial superconducting transitions, or resistive upturns below $T_\textrm{c}$ (Figures S24--S51). Recently, a transition from superconducting to insulating behavior was reported in fully amine-intercalated 2$M$-WS$_2$, attributed to reduced interlayer coupling.~\cite{Xie2026} Our results suggest that the degree of layer coupling and corresponding electronic behavior may be tunable, rather than restricted to discrete superconducting or insulating states. Controlled exchange of interstitial ions or neutral polar molecules could offer a rational route to tune superconductivity in layered AMID materials. 
\par
The structure resulting from each deintercalation treatment also varies with the amount of residual K$^+$. Prior EDS and XPS measurements on deintercalated samples have been interpreted as evidence that K$^+$ is removed completely~\cite{Fang2019,Lai2021,Zhou2026}. Both techniques have detection limits of $\sim$1,000 ppm under ideal conditions. Residual K$^+$ present at concentrations near or below these detection limits may go unobserved. We quantified residual K$^+$ by inductively coupled plasma mass spectrometry (ICP-MS) (for experimental details and discussion of uncertainty see the Supporting Information). ICP-MS can provide accurate concentrations for samples containing below one ppb of K$^+$. Using this technique, we found a significant and measurable amount of residual K$^+$ intercalant in all samples tested (Figure \ref{ICP}).
\\ 
\indent
The amount of residual K$^+$ appears to impose an upper bound on crystal quality and could reflect the equilibrium position of the deintercalation reaction. This observation is consistent with our $\overline{RRR}$ results and suggests that more complete deintercalation tends to yield higher-quality crystals. While crystal quality generally improved with decreasing residual K$^+$, multiple conditions that achieved near-complete K$^+$ removal yielded crystals with very low $\overline{RRR}$. In these cases, damage to the lattice (e.g., 0.01 M H$_2$SO$_4$) or transformation of the crystal structure during treatment (e.g., 1\% bleach) may account for samples falling below this bound. For treatments yielding crystals with low residual K$^+$ and high $\overline{RRR}$, residual K$^+$ likely no longer limits crystal quality. Further improvement of $\overline{RRR}$ beyond the highest values achieved here presumably requires suppressing unproductive side reactions during treatment or reducing defects in the WS$_2$ layers that form during the initial growth of 2$M$-K$_x$WS$_2$.  
\FloatBarrier
\section{Conclusion}
Our work suggests that 2$M$-K$_x$WS$_2$ deintercalation is not a single oxidation event but is instead an equilibrium operating within a network of coupled and competing reactions. As expected for an equilibrium, deintercalation of K$^+$ was incomplete under every condition tested. Crystal quality does not appear to be conferred by the oxidative mildness of the treatment but instead by the position of the equilibrium and suppression of damaging side reactions. Synthesis in solid-state chemistry is often practiced as a static exercise, in which well-defined reactants are converted into supposedly well-characterized products via opaque reaction processes.~\cite{Schollhorn1996} Framing the deintercalation of 2$M$-K$_x$WS$_2$ as a chemical reaction with a testable mechanism could transform it from an empirical treatment into a rational synthetic method whose conditions can be tuned to produce higher-quality crystals. Based on our results, it seems likely that deintercalation reactions in other AMID systems could be thermodynamically controlled. If so, the mechanistic framework proposed here could guide the synthesis of higher-quality crystals or entirely new phases within these systems, both of which are likely required for observing new physics, such as topological superconductivity, and for rationally designing functional materials with specific properties.

\begin{acknowledgement}
 We are grateful to Philip P. Lampkin (University of Utah) for many insightful discussions. We also thank Dr. Ilia A. Guzei (UW–Madison) for assistance with SCXRD data collection and analysis. This work was primarily supported by the National Science Foundation under grant No. DMR-2338984 and the University of Wisconsin–Madison. The authors gratefully acknowledge the use of facilities and instrumentation in the Wisconsin Center for Nanoscale Technology. This center is partially supported by the Wisconsin Materials Research Science and Engineering Center (NSF DMR-2309000) and by the University of Wisconsin–Madison. The Quantum Design MPMS3 magnetometer was supported by the University of Wisconsin-Madison Department of Chemistry. The purchase of the Bruker D8 VENTURE Photon III Mo Diamond II I$\mu$S X-ray diffractometer was partially funded by a 2024 University of Wisconsin Research Core Revitalization Program Round 3 Award to the Department of Chemistry. ICP-MS was performed at the Core Facility for Advanced Water Analysis at the Water Science and Engineering Laboratory (WSEL Core), University of Wisconsin–Madison.
\end{acknowledgement}

\begin{suppinfo}
Experimental and instrumentation details, including additional discussion of crystal structures, supplementary figures, and tables (PDF)

Deposition numbers 2593082 and 2593083 contain the supplementary crystallographic data for this paper. These data can be obtained free of charge via the joint Cambridge Crystallographic Data Centre (CCDC) and Fachinformationszentrum Karlsruhe Access Structures service.
\end{suppinfo}


\begin{mcitethebibliography}{77}
\bibitem[Not()]{Note-1}
E$_T$(30) is defined as the molar electronic transition energy (in kcal/mol) of
  the negatively solvatochromic betaine dye 30, calculated from its
  longest-wavelength charge-transfer absorption.\relax
\mciteBstWouldAddEndPunctfalse
\mciteSetBstMidEndSepPunct{\mcitedefaultmidpunct}
{}{\mcitedefaultseppunct}\relax
\EndOfBibitem
\end{mcitethebibliography}
\providecommand{\latin}[1]{#1}
\makeatletter
\providecommand{\doi}
  {\begingroup\let\do\@makeother\dospecials
  \catcode`\{=1 \catcode`\}=2 \doi@aux}
\providecommand{\doi@aux}[1]{\endgroup\texttt{#1}}
\makeatother
\providecommand*\mcitethebibliography{\thebibliography}
\csname @ifundefined\endcsname{endmcitethebibliography}
  {\let\endmcitethebibliography\endthebibliography}{}

\end{document}